\documentclass[pra,twocolumn,amsmath,amssymb,groupedaddress,superscriptaddress]{revtex4-1}
\usepackage{graphics}
\usepackage{amsmath}
\usepackage[dvips]{graphicx}
\usepackage{float,graphicx}
\usepackage{epsfig}
\usepackage{epstopdf}

\def\>{\rangle}
\def\<{\langle}

\def\be{\begin{equation}}

\def\ee{\end{equation}}

\def\eeff{\textmd{eff}}

\def\iint{\textmd{int}}

\def\be{\begin{equation}}
\def\ee{\end{equation}}
\def\bea{\begin{eqnarray}}
\def\eea{\end{eqnarray}}
\usepackage{bm}
\usepackage{graphicx}

\begin{document}

\title{Quantum dynamics in photonic crystals}

\author{Javier Prior}
\affiliation{Departamento de F\'isica Aplicada, Universidad Polit\'ecnica de Cartagena, Cartagena 30202, Spain}
\author{In\'es de Vega}
\affiliation{Institut f\"{u}r Theoretische Physik, Albert-Einstein-Allee 11, Universit\"{a}t Ulm, D-89069 Ulm, Germany}
\author{Alex W. Chin}
\affiliation{Institut f\"{u}r Theoretische Physik, Albert-Einstein-Allee 11, Universit\"{a}t Ulm, D-89069 Ulm, Germany}
\affiliation{Theory of Condensed Matter Group, University of Cambridge,
J J Thomson Avenue, Cambridge, CB3 0HE, United Kingdom}
\author{Susana F. Huelga}
\affiliation{Institut f\"{u}r Theoretische Physik, Albert-Einstein-Allee 11, Universit\"{a}t Ulm, D-89069 Ulm, Germany}
\author{Martin B. Plenio}
\affiliation{Institut f\"{u}r Theoretische Physik, Albert-Einstein-Allee 11, Universit\"{a}t Ulm, D-89069 Ulm, Germany}

\begin{abstract}

Employing a recently developed method that is numerically
accurate within a model space simulating the real-time dynamics of few-body systems interacting with macroscopic environmental quantum fields, we analyze the full dynamics of an atomic system coupled
to a continuum light-field with a gapped spectral density. This is a situation encountered, for
example, in the radiation field in a photonic crystal, whose analysis has been so
far been confined to limiting cases due to the lack of suitable numerical techniques. We
show that both atomic population and coherence dynamics can drastically deviate from
the results predicted when using the rotating wave approximation, particularly in the
strong coupling regime. Experimental conditions required to observe these corrections
are also discussed.
\end{abstract}

\date{\today}

\maketitle
\section{Introduction}
In the past decades, there has been an enormous interest in developing methods to
control and modify the light field. In most cases, this has been achieved by building specific materials where the light matter interaction strongly modifies the characteristics of the free electromagnetic field. These new materials result in interesting technological applications as well as theoretical challenges. Important examples are microcavities \cite{vahala2003}, and metamaterials \cite{Pendry1999}, which are designed respectively to strengthen the light-matter interaction, and to tailor the refraction index altering the flow of light in a non-trivial way \cite{pendry2006,tsakmakidis2007}.

Within this scenario, photonic crystals (PC) provide one of the most interesting examples of artificially engineered materials. PCs \cite{yablonovitch1987,john1987,john1995} (see \cite{woldeyohannes2003,angelakis2004} for basic reviews) are periodically microstructured compounds which tailor the vacuum electromagnetic density of states, creating specific frequency ranges, or gaps, where it vanishes. Reduction or suppression of the density of states within the band gap facilitates light localization and trapping in a bulk material \cite{yablonovitch1987,john1987,baba2001}, as well as the inhibition of spontaneous emission over a broad frequency range. Additionally, the density of states varies rapidly near the edge of the gap, whose frequency we will denote by $\omega_b$, which implies that the correlation time of the vacuum fluctuations can be comparable to the characteristic time scales of atoms, quantum dots or NV centers  embedded in the PC structure. Hence, an accurate description of the dynamics of systems in contact with highly structured environments, like the radiation field within PCs, may force us to go beyond the usual Markovian, and weak coupling approximations \cite{breuerbook,rivas2010,rivas2011}. Therefore, a reliable method to describe the full system dynamics is highly desirable.

Here, we demonstrate that the efficient, numerically accurate method developed in \cite{chin2010,prior2010,chin2011} (TEDOPA: Time Evolving Density with Orthogonal Polynomials Algorithm) is an excellent candidate to describe the dynamics of quantum systems, such as impurity atoms and quantum dots, within structured reservoirs and quantum fields. To this end, we apply TEDOPA to describe the dynamics of a two level atom embedded in a photonic crystal. Our reason to chose such a system as a test bed is that, under certain limits and within the rotating wave approximation (RWA), it is known to be exactly solvable, thus providing an excellent arena to prove the reliability of our technique.
Most importantly, the method deployed here is valuable in describing the dynamics without the usual weak coupling and RWA. By evolving the system without invoking RWA, strong departures from the RWA result are found. We will show that
when the system-environment coupling is strong, the atomic population dynamics differ from the RWA solution and an enhanced population trapping can be observed. When the system energy splitting, $\Delta$, is much smaller than any other system and environment time scale, the RWA leads to an unphysical prediction for the system's coherence in the low frequency domain while TEDOPA allows to evaluate the accurate dynamics across the whole frequency domain.

\section{The model}\label{model}

We consider a single atom or quantum dot, modeled as a two-level system with transition frequency $\Delta$, coupled to the modified radiation field that exists within a photonic crystal. The general interaction Hamiltonian can then be written in interaction picture as follows ($\hbar=1$) \cite{woldeyohannes2003,devega2005, devega2008}:
\begin{eqnarray}
H_{\iint}&=&\sum_{\textbf{k},\mu}g_{\textbf{k},\mu} \bigg(
b^\dagger_{\textbf{k},\mu} \sigma^- e^{i \Delta^{-}_{{\bf k}} t}
+b^\dagger_{\textbf{k},\mu}\sigma^+e^{i \Delta^{+}_{{\bf k}} t}+h.c.\bigg), \label{genH2}
\end{eqnarray}
where $\sigma^{\pm}$ are the atomic spin ladder operators, $\mu=1,2$ are the two polarization modes of the light, and ${\bf k}$ is the field wave vector. In writing the Hamiltonian in this way, we ignore, for simplicity, any effects related to light polarisation and the existence of multiple bands in photonic cystrals \cite{john1995,woldeyohannes2003,devega2005}. The frequencies are defined as $\Delta^{\pm}_{{\bf k}}=\omega_k\pm \Delta$, and we consider $\omega_k=\omega_b+k^2/2m$, where $m$ is an effective mass acquired by a photon in a PC \cite{john1995,woldeyohannes2003},  corresponding to the dispersion relation of the radiation field for frequencies in the vicinity of a single band-gap edge with frequency $\omega_b$, and $k=|{\bf k}|$. Following Refs. \cite{wallsbook,devega2005}, the microscopic form for the coupling between the light field and optical transition is given by $g_{{\bf k},\mu}=\Omega_{k,\mu} e^{-X_{0}^2 k^2 /2}$, with $\Omega_{k,\mu}=\Delta\sqrt{\frac{1}{2\hbar\omega_k\epsilon_0 v}}d_{12} {\bf \hat{\mathbf{d}}}\cdot{\bf \epsilon_{k,\mu}}$, where $\epsilon_0$ is the electric permittivity in vacuum, $v$ is the quantization volume, ${\bf \hat{\mathbf{d}}}$ and $d_{12}=X_0^2 k e/2$ are respectively the direction and magnitude of the dipole moment with $e$ the electric charge and $X_0$ the Bohr radius, and ${\bf \epsilon_{k,\mu}}$ for $\mu=1,2$ represents the unitary polarization vector corresponding to the two transverse polarization modes. The factor $e^{-k^2 X_0^2/2}$ physically arises from the fact that when calculating the coupling (see for instance \cite{wallsbook}), one should take into account that the electronic wave functions have a finite width $X_0$. Here, we have considered that the electronic wave function for the electronic level $j$ is approximately given by $\phi_j ({\bf r})\approx \frac{1}{\pi^{3/2}X_0^{3/2}}e^{-({\bf r}-{\bf r}_j)^2/(2X_0^2)}$. In other words, the factor $e^{-k^2 X_0^2/2}$ appearing in $g_{k,\mu}$, often neglected within the dipolar approximation, is due to the fact that the electronic wave function varies on a scale that is approximately determined by the Bohr radius, $X_0$ \cite{wallsbook}. Considering that the electron moves in a finite space leads to an effective momentum cut-off for the emitted photons, $k_0 =2\pi/X_0$, and thus introduces a maximum photonic frequency $\omega_0=\omega_{k_0}$ into the problem.

The bosonic creation and annihilation operators of the photonic crystal light modes are $b_{{\bf k},\mu}$ and $b_{{\bf k},\mu}^{\dagger}$, respectively for a mode of wavenumber $k$ and polarization $\mu$.
The above Hamiltonian is not exactly solvable and the RWA is usually considered, so that the fast rotating terms (i.e. terms $\sim b_{{\bf k},\mu}\sigma^-$ or $b_{{\bf k},\mu}^\dagger\sigma^+$), are neglected with respect to the other (energy conserving) terms. In this case, $H_{\iint}^{RWA}= \sum_{{\bf k},\mu}g_{{\bf k},\mu} \left(b^\dagger_{\bf
k,\mu}\sigma^- e^{i \Delta^-_{\bf k} t}+h.c.\right)$.
While the RWA is often a very good approximation, due to the existence of well separated time-scales, this is not always the case. In the case of strong coupling, we shall show that large deviations from RWA physics emerge, and that their description requires sophisticated numerical techniques.

\section{Single atom dynamics with the RWA}\label{single}

A single atom constitutes the simplest possible case, but represents a valuable example to prove the validity or our method.
As noted above, when considering the RWA, the dynamics specified by the Hamiltonian Eq.(\ref{genH2}) always remains in the one excitation sector and the time-dependent Schr{\"o}dinger equation for the total atom-photonic crystal system wave function can be solved exactly \cite{woldeyohannes2003,devega2008}. In this regard, the wave function of the total system can be written as,
\begin{equation}
|\Psi(t)\rangle=A(t)|1,\{0_{\mathbf{k}}\}\rangle+\sum_{\bf k,\mu}B_{\bf
k,\mu}(t)|0,{1}_{\bf k,\mu}\rangle,\end{equation}
where $|1,\{0_{\mathbf{k}}\}\rangle$ describes the atom excited and no photon in the radiation field, and
$|0,{1}_{\bf k,\mu}\rangle$ represents no excitation in the atom, and a single photon in the mode ${\bf k,\mu}$. Note that this form for the system wave function is valid only when the RWA is assumed, and the dynamics is restricted to the one excitation sector.

The time-dependent Schr{\"o}dinger equation projected on the one-atom sector of the Hilbert space takes the form,
\begin{eqnarray}
\frac{dA(t)}{dt}&=&-\sum_{\bf k,\mu}g_{\bf k,\mu} B_{\bf k,\mu}(t)e^{-i \Delta_{\bf k}^{-}t},\\ \frac{dB_{\bf k,\mu}(t)}{dt}&=&g_{\bf k,\mu} A(t)e^{i\Delta_{\bf k}^{-}t}.\end{eqnarray}
Inserting the formal solution of the second equation into the first one, we have
\begin{eqnarray}
\frac{dA(t)}{dt}=-\int_0^t d\tau G(t-\tau)A(\tau),\label{oneatomsector3}
\end{eqnarray}
where
\begin{eqnarray}
G(t)&=&\sum_{\bf k,\mu} g_{\bf k,\mu}^2 e^{-i\Delta^-_{\bf
k}t}
\nonumber\\
&=&\frac{4\pi X^3_0 \Omega^2}{\pi^{3/2}}\int_{1BZ} dk k^2 e^{-i(\frac{k^2}{2m}+\omega_b-\Delta)t}e^{-X_0^2k^2}\nonumber\\
&=& \Omega^2\frac{e^{i (\Delta-\omega_b) t}}{(1+i\omega_0 t)^{3/2}}
\end{eqnarray}
is the correlation function of the system-environment coupling \cite{devega2008,navarrete2010}.
We note that in order to calculate the correlation function, we have assumed that the factor $|{\bf\hat{d}}\cdot{\bf \epsilon_{k,\sigma}}|^2/\omega_k$ changes very smoothly with ${\bf k}$ in the 1BZ (first Brillouin zone) of the photonic crystal, and can be considered a constant factor in comparison with the rapidly oscillating exponential factor \cite{devega2005}. As discussed in \cite{devega2005} the validity of this approximation can be easily verified numerically, and it allows us to settle $\sum_{\mu=1,2}\Omega_{k,\mu}^2\approx \Omega^2=2\frac{\Delta^2d_{12}^2}{2\hbar\omega_b\epsilon_0 v}$. Here, we have considered in addition that the sum over the two polarization modes gives rise to a factor two.
Note that the high-frequency cut-off arising from the finite size of the electronic wave functions is required to avoid a singularity at the origin of time in the correlation function (such singularity appears for instance in \cite{florescu2004}).

A change of variable between $k$ and $\omega_k$, leads to expressing the correlation function as
 $G(t)=\frac{1}{\pi}\int_0^\infty d\omega J(\omega)e^{-i(\omega-\Delta) t}$,
with the spectral density given by
\begin{eqnarray}
J(\omega)&=&\frac{1}{\pi}\sum_{{\bf k},\mu}g_{\textbf{k},\mu}^2\delta(\omega-\omega_{\textbf{k}})\nonumber\\
&=&\alpha\sqrt{\omega-\omega_b}e^{-\frac{\omega-\omega_b}{\omega_0}},
\label{spectralgap}
\end{eqnarray}
where $\alpha=2\pi\Omega^2/(\omega_0^{3/2})$. The amplitude $A(t)$ can be obtained by taking the Laplace transform of Eq. (\ref{oneatomsector3}) and inverting the resulting expression for the Laplace transform $\mathcal{A}(s)$ of $A(t)$. From Eq. (\ref{oneatomsector3}) we thus need to invert $\mathcal{A}(s)=A(0)/(s+\mathcal{G}(s))$, where $\mathcal{G}(s)$ is the Laplace transform of $G(t)$. Once determined, the population of the atom in the excited state at time $t$ is given by $|A(t)|^2$.

Using the Laplace transform method, a numerical solution for $A(t)$ can always be obtained. In contrast, an analytical solution requires additional approximations, namely that the system frequencies are very small compared to the cut-off frequency $\omega_0$, so that $\omega_0 \gg \Omega,\Delta,\omega_b$, and that all relevant time scales obey $t\gg1/\omega_0$ \cite{john1994,john1995}. Under these conditions, this leads to $G_\infty(t)=-\alpha e^{i(\Delta-\omega_b)
t+\pi/4}/t^{3/2}$, which is singular at the origin, but describes
correctly times $t\gg 1/\omega_0$ \cite{john1994,john1995,devega2008}.
Inverting the Laplace transform of $\mathcal{A}(s)$ with this approximation for $G(t)$, the following solution is obtained \cite{devega2008}

\begin{equation}
A(t)=c_1 e^{i (r_1^2 +\Delta_L)t}+I(\alpha,\Delta_L,t),\label{sol}
\end{equation}
with $I(\alpha,\Delta_L,t)=(\alpha /\pi) \int_0^\infty dx
\frac{\sqrt{x}e^{(-x+i \Delta_L)t} }{(-x+i \tilde{\Delta}^-_L)^2+i \alpha^2 x}$. Defining $r_{\pm}=-(\alpha/2)\pm\sqrt{(\alpha/2)^2-\Delta_L}$ and $\Delta_L=\Delta-\omega_b$, we have three different regimes:
(i) If $\alpha^2/2> \tilde{\Delta}_L >0$, then $c_1=0$, (ii) If $\tilde{\Delta}_L>\alpha^2/2$, then $r_1=r_-$ and $c_{1}=\frac{2 r_-}{r_--r_+}$, and finally, (iii) if $\tilde{\Delta}_L <0$, $r_1=r_+$ and $c_1=\frac{2r_+}{r_+ -r_-}$.
Under these conditions, several dynamical regimes exist depending on the value of the parameter $\tilde{\Delta}_L=\Delta_L-\omega_s$, where $\omega_s$ is a frequency that arises from renormalisation of the optical transition by the environment, $\omega_s=4\Omega^2/\omega_0$. Like the Lamb-shift encountered in traitional master equation approaches, $\omega_{s}$ depends on the coupling strength and is discussed in Refs. \cite{devega2008,navarrete2010} and in Section \ref{coherences}. In this regard, for $\tilde{\Delta}_L>0$ the atom is not excited in the stationary state (i.e. relaxes completely to the ground state), whereas for negative values of $\tilde{\Delta}_L$ there is a stationary residual excited atomic population (corresponding to a situation where the emitted photon remains localized nearby the atom). Thus, tuning $\tilde{\Delta}_L$ from negative to positive values leads to a cross-over between two distinct regimes \cite{devega2008,navarrete2010}.

\subsection{Chain mapping, rotating wave dynamics and TEDOPA}

When the RWA is not applicable, the dynamics of the global atom-light system cannot be described in the one-excitation manifold, and the direct wave function approach presented above becomes intractable. However, a new technique has recently emerged in the theory of open quantum systems which allow this task to be performed with numerical exactitude. This technique \cite{prior2010,chin2010} employs time-adaptive density matrix re-normalisation group (t-DMRG) algorithm to compute the evolution of the full atom and light field wavefunction. The t-DMRG algorithm enables extremely accurate simulation of large many-body systems with nearest-neighbour interactions in $1D$, and has become widely used in cold atom and condensed matter theory \cite{scholl2011}.

To apply this technique to the present problem, we start with the full atom-light field Hamiltonian in the continuum Schr{\" o}dinger picture,
\begin{eqnarray}
H&=&\frac{\Delta}{2} (1+\sigma_{z})+\sigma_{x}\int_{0}^{1}dkh(k)\,(b(k)+b(k)^{\dagger})\nonumber\\
&+&\int_{0}^{1}\omega(k)b(k)^{\dagger}b(k)
\label{Hamil3}
\end{eqnarray}
where the field operators obey that $[b(k),b(k')^{\dagger}]=\delta(k-k')$, $h(k)$ is the atom-light coupling strength and $\omega(k)$ is the dispersion of the photons. The parameter $k$ labels the modes and is normalised so that the highest frequency of the modes, that we will denote by  $\omega_{c}$, occurs at $k=1$. When the environment is initially in a gaussian state, it is possible to integrate out the environment degrees of freedom and derive a formally-exact expression for the reduced state dynamics of the atomic system using path integral methods \cite{leggett1987}. Importantly for the application of t-DMRG, this result shows that the effects of the photonic environment will only enter through the spectral function $J(\omega)$ of the photons \cite{leggett1987}. In the following simulations, we assume that the initial state of the total atom and environment is a product state of an atomic state and the zero-temperature vacuum state of the light field.  As we are only interested in the reduced state of the atom, we can make use of the fact that the only relevant environmental property is the spectral function to represent the bosonic environment in a form that allows us to transform it into an equivalent system suitable for TEDOPA simulation.

In terms of the continuum dispersion and coupling functions, the spectral function defined in Eq. (\ref{spectralgap}) takes the general form $J(\omega)=h^{2}(\epsilon(\omega))\frac{d \epsilon (\omega)}{d\omega}$, where $\epsilon(k)$ is the inverse function of the dispersion $\omega(k)$, i.e. $\epsilon(\omega(x))=x$ \cite{chin2010}.  From this one can see that a variety of different pairs of $h(k)$ and $\omega(k)$ can lead to the same spectral function, and we exploit this. Taking $h(k)^2=\omega_{c}J(\omega(k))$ and the linear dispersion $\omega(k)=\omega_{b}+\omega_{c}k$, with $\omega_c$ the maximum frequency of the environment considered in the simulations, it can be found that the spectral function characteristic of (\ref{Hamil3}) (as described in \cite{chin2010}) is given by $J(\omega)\Theta(\omega_{c}-\omega)$, where $J(\omega)$ is the spectral function of Eq.(\ref{spectralgap}) and $\Theta(x)$ is the Heaviside function. As the dynamics of the reduced atomic state only depend on $J(\omega)$, the t-DMRG simulations (with these choices of $\omega(k)$ and $h(k)$) simulate exactly the same reduced dynamics implied by Eq. (\ref{genH2}) with the physcial microscopic dispersions and coupling functions described in Section \ref{model}. Because of the hard cut-off in the spectral function, dynamics are captured accurately on all frequency scales up to $\omega_{c}$, which is chosen to be much larger than the physcial scale $\omega_{0}$ that determines the frequency range over which $J(\omega)$ is significantly different from zero. Therefore, by taking $\omega_{c}\gg\omega_{0}$ the t-DMRG describes all relevant physics beyond times of $\approx 1/\omega_{c}$ with numerical exactitude.

We then perform a unitary transformation of the Hamiltonian given in Eq. (\ref{Hamil3}) by  defining new environment modes $a_{n}=\rho_{n}^{-1}\int_{0}^{1}h(k)\pi_{n}(k)b(k)$, where $\pi_{n}(k)$ are monic orthogonal polynomials which obey $\int_{0}^{1} dk\,J(g(k))\pi_{n}(k)\pi_{m}(k)=\rho_{n}^{2}\delta_{nm}$.  By choosing a linear-in-k dispersion, the properties of the monic orthogonal polynomials ensure that this transformation is both real orthogonal and leads to a nearest-neighbour chain Hamiltonian given by \cite{chin2010,chinbook2011},
\begin{eqnarray}
H_{\textmd chain}&=&\frac{\Delta}{2} (1+\sigma_{z})+ g\sigma_{x}(a_{0}+a_{0}^{\dagger})\\
&+&\sum_{n=0}^{\infty}(\omega_{b}+\epsilon_{n})a_{n}^{\dagger}a_{n}+t_{n}a_{n}^{\dagger}a_{n+1}+t_{n}a_{n+1}^{\dagger}a_{n}.\nonumber
\label{map}
\end{eqnarray}
Within this picture, the two level atom now couples to one end of a chain of coupled harmonic oscillators. The frequencies $\epsilon_{n}$ and couplings $t_{n}$ are simply determined by the recurrence coefficients of the monic orthogonal polynomials which are effectively determined by the spectral function (in the language of orthogonal polynomials, the spectral function is treated as a weight function)  \cite{chin2010,chinbook2011}.  The use of the theory of orthogonal polynomials also allows rigourous proof of this mapping for almost any physical spectral function, including continuous, discrete, mixed and gapped spectra \cite{chin2010,chinbook2011}. We also note more generally that this procedure effectively enables a highly efficient way of simulating one-dimensional quantum field dynamics coupled to non-linear objects, such as spins, and also other quantum fields.

After this procedure it now becomes possible to apply time-adaptive density matrix renormalisation methods \cite{vidal2003,scholl2011}
to simulate the unitary evolution of the total atom and chain wavefunction $|\Psi (t)\rangle$. From $|\Psi (t)\rangle$, the real-time expectation of any observable $O_{a}$ of the atom can be simply calculated by evaluating $\langle O_{a}(t)\rangle=\langle\Psi(t)|O_{a}|\Psi(t)\rangle$. Although we will not pursue this here, another advantage of this method is that we also have full access to the  evolving state of the environment field itself. This can provide deep insight into real-time dissipative processes, as has been recently demonstrated in a study of exciton transport in photosynthetic complexes \cite{chin2012}. The combination of the powerful analtyical mapping with the numerical strength of t-DMRG we shall henceforth refer to as the TEDOPA (Time Evolving Density with Orthogonal Polynomials Algorithm) technique, and a comprehensive overview of its theory, implementation and applications can be found in \cite{chinbook2011}.

\subsection{ A quick comparison of methods}
While numerical simulations with t-DMRG can be made arbitrarily precise in principle, in practice finite bond-dimension employed in the simulation will limit precision. The numerical simulation error can be bounded from above but the resulting bounds are very conservative as they are growing in time. Hence it is instructive  to compare simulation results against know exact solutions.

\begin{figure}[ht]
\centerline{\includegraphics[width=0.38\textwidth]{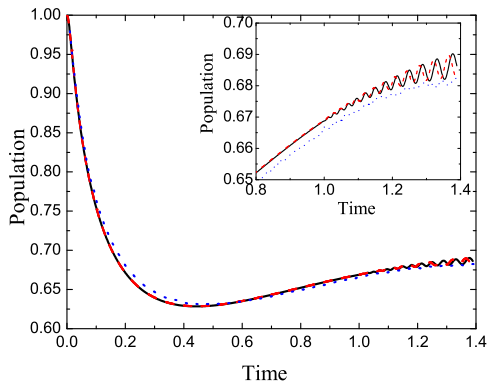}}
\caption{ (Color online) Comparison between t-DMRG (solid black), Heisenberg (dashed red), and Piessens (blue dots) solutions for $|A(t)|^2$, for the parameters $\alpha=1$, $\omega_{0}=100$, gap-edge frequency $\omega_{b}=5$, $\omega_{c}=800$ and $\Delta=1$. Time is in units of $1/\alpha^2$. \label{fig1}}
\end{figure}

To this end, we consider the RWA approximation and compare the results obtained using TEDOPA and the numerical solution of $A(t)$. As the analytical (inverse Laplace transform) solution for $A(t)$ discussed in the previous section is only valid when $t\gg1/\omega_0$ and $\omega_0 \gg \Omega,\Delta,\omega_b$, we use the Piessens method \cite{piessens73} to numerically invert the Laplace transform of $A(t)$ for the large but finite values of $\omega_{0}$ used in our TEDOPA calculations. In addition, it is possible to numerically integrate the RWA Heisenberg equations of motion in the chain representation of Eq. (\ref{map}) to obtain a numerically accurate evolution of the atomic population. Figure \ref{fig1} shows the population ($|A(t)|^2$) in the excited state of the atom for $\alpha=1$, $\omega_{0}=100$, gap-edge frequency $\omega_{b}=5$, $\omega_{c}=800$ and $\Delta=1$ calculated by t-DMRG, the Piessens method and the numerical integration of the Heisenberg equations.  The figure shows that the t-DMRG result converges to both exact RWA results, while the inset shows that the finite system-size recurrence features which appear in the Heisenberg integration scheme are reproduced almost exactly if the same number of sites are used in the t-DMRG simulation. We note that it is also possible to perform t-DMRG in the Heisenberg picture, and that for the RWA example considered here it will be numerically accuarate \cite{hartman09,clark10}. Anticipating later results, which show that TEDOPA captures the non-rotating wave physics extremely well, we might also expect Heisenberg t-DMRG to perform well in the non-rotating wave case \cite{hartman09,clark10}.  The slight departure between the t-DMRG and Heisenberg method from the Piessens method is due to the growing inaccuracy and instability of the Piessens method (a general feature of many numerical Laplace inversion techniques) as time evolves \cite{piessens73}.

\section{Accurate Numerical Results}
\begin{figure}[ht]
\centerline{\includegraphics[width=0.4\textwidth]{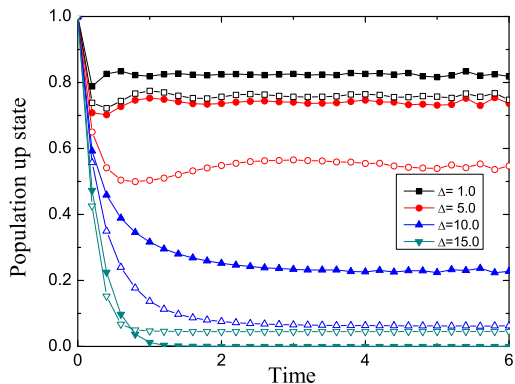}}
\caption{ (Color online) Excited state population dynamics for the RWA (filled plot markers) and full Hamiltonian (empty plot markers).  For these data $\alpha=1$,$\omega_{0}=100$, $\omega_{b}=5$, and $\omega_{c}=800$.  Time is in units of $1/\alpha^2$\label{fig2}}
\end{figure}

\begin{figure}[ht]
\centerline{\includegraphics[width=0.4\textwidth]{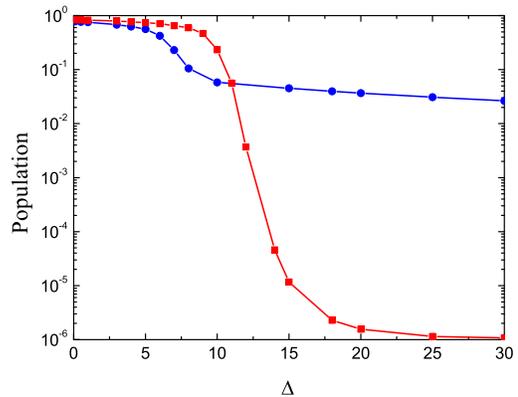}}
\caption{ (Color online) Stationary state atomic population for the RWA (red squares) and the complete Hamiltonian (blue dots) as a function of $\Delta$. The initial state is a fully excited system, and we have used $\alpha=1$, $\omega_{0}=100$, $\omega_{b}=5$, and $\omega_{c}=800$. \label{fig3}}
\end{figure}

The cross-over occurring when tuning $\tilde{\Delta}_L$ from negative to positive values can also be observed when considering the exact numerical solution of $A(t)$, but until now it had only been described within the RWA.  Thus, a question that remained open is whether this transition is a consequence of the RWA, or can also be observed in situations where the RWA is invalid. To explore this we now present results in the regime of strong coupling where we expect, and shall demonstrate, that deviations between RWA and the full simulations to be most apparent.  This regime of strong coupling is reached when the energy shift induced by the environment coupling when $\Delta=0$ is comparable to the energy splitting of the optical transition when $\Delta$ is finite. The environmental energy (polaron) shift is $E_{en}=\pi^{-1}\int_{\omega_{b}}^{\infty}J(\omega)\omega^{-1}\approx \alpha(\omega_{0}/\pi)^{\frac{1}{2}}$ \cite{chin2011}, and by taking parameters $\alpha=1$, $\omega_{0}=100$, $\omega_{b}=5$ in the rest of this article, we ensure that we remain between the strong and intermediate coupling regimes for all of the values of $\Delta$ we shall consider.

Fig. \ref{fig2} shows the time evolution of the population in the excited state as a function of time, when starting from the excited atomic state, and for fixed bath coupling and variable $\Delta$. Clear differences in the evolution can be seen, particularly in the final values of the excited state population. Indeed, when $\Delta=15$ the result of the RWA predicts no population in the excited state, whereas the dynamics for the full Hamiltonian (\ref{genH2}) still shows a non-zero final excited state population. In general, the RWA overestimates the excited state stationary population for $\tilde{\Delta}_{L}<0$, and underestimates this quantity for $\tilde{\Delta}_{L}>0$, as seen for $\Delta=10$. We also note the oscillatory features of the population dynamics, which in both cases decrease in frequency as $\Delta$ approaches the band edge. However, at fixed $\Delta$ the RWA and full Hamiltonian frequencies are markedly different, especially when $\Delta=\omega_{b}$. This behaviour will be rationalised in section \ref{coherences}.

Fig. \ref{fig3} shows the stationary atomic population for an initially excited atom, with respect to the atomic frequency $\Delta$. In this Figure, the mentioned cross-over from a photon-atom bound state to a total relaxed state is sharper within the RWA than when the full Hamiltonian Eq. (\ref{genH2}) is simulated. A clear difference in the residual populations above the cross over is also observed. As we shall see, these differences arise due to the improper treatment of the renormalisation of the atomic transition energy in the RWA. However, at very weak coupling, these differences are negligible and the cross-over happens at $\Delta\approx\omega_{b}$ (not shown). We also note that a sharp `transition' at $\tilde{\Delta}_L=0$ only appears as $\omega_{0}\rightarrow \infty$ \cite{woldeyohannes2003,devega2008}, and when the coupling is very weak. However one should also note that the cross-over width in Fig. \ref{fig3} is over-emphasised by the logarithmic scale.

Fig.\ref{fig3} also shows that the residual population in the excited state remains larger in the simulations of the full Hamiltonian relative to results obtained within the RWA approximation. This arises due to \emph{multi-photon} effects which cannot be described in RWA. For $\Delta\gg \omega_{b}$, the atom is able to exchange energy with the light field via photon emission/absorption, and at long times relaxes towards the ground state of the \emph{ full} atom-photon Hamiltonian. However, the interactions with the bath alter the effective eigenstates of the system, leading to the residual excited state populations in the atom-light field ground state. This state can be approximately described by the variational polaron ground state, first suggested in \cite{silbey1984}. Following the procedure set out in \cite{silbey1984}, the reduced density matrix of the atom's variational polaron ground state - in the basis $|\pm\rangle$, where $|\pm\rangle$ are the eigenstates of $\sigma_{x}$ - is given by $
\rho_{gs}^{atom}=\frac{1}{2}|+\rangle\langle +|+\frac{1}{2}|-\rangle\langle-|-\Phi|-\rangle\langle +|-\Phi|+\rangle\langle -|$ where $\Phi=\tilde{\Delta}/\Delta$ and $\tilde{\Delta}$ is determined from the implicit equation \begin{equation}
\tilde{\Delta}=\Delta\exp\left[-\frac{2}{\pi}\int_{\omega_{b}}^{\infty}\,\frac{d\omega J(\omega)}{(\omega+\tilde{\Delta})^2}\right].\label{renorm}\end{equation}

The off-diagonal elements of the density matrix are suppressed by dressing, or polaronic, correlations (state-dependent coherent displacements) between the bath and atomic states $|\pm\rangle$. These correlations effectively suppress the effective energy gap ($\tilde{\Delta}\rightarrow 0$) between the atomic states $|\uparrow\rangle$ and $|\downarrow\rangle$ through the reduced overlap of the displaced light mode wave functions which dress the states $|\pm\rangle$ \cite{silbey1984,chin2011}. From the density matrix $\rho_{gs}^{atom}$, the population in the excited atomic level is $P_{\uparrow}= \frac{1}{2}(1-\Phi)$. For large $\alpha,\omega_{b}\ll \Delta \ll \omega_{0}$ the approximate self-consistent solution of Eq. (\ref{renorm}) is $\tilde{\Delta}\approx\Delta \left(1-\frac{\alpha}{\sqrt{\Delta}}\right)$. This shows that as $\Delta$ gets larger, the re-normalisation from the environment gets smaller, so $\Phi\rightarrow1$ and $P_{\uparrow}\rightarrow 0$. For the parameters in Fig. \ref{fig3}, we find that the solution of Eq. (\ref{renorm}) with $\Delta=30$ gives $P_{\uparrow}=0.026$ which agrees with the data very well. Note that these formulas are only valid when the atom is able to relax to the polaronic ground state, which occurs for  $\Delta >\omega_b$ \cite{shift}; at $\Delta< \omega_{b}$, relaxation is blocked by energy conservation. In this case a better prediction for the residual population is to assume that the final state is the dressed excited state, in which case the final population is $P_{\uparrow}= \frac{1}{2}(1+\Phi)$, with $\Phi$ obtained from Eq. (\ref{renorm}).  In the RWA approximation the analytical theory predicts a total de-excitation of the atom at large $\Delta > \omega_{b}$. This is seen in Fig. \ref{fig3}, although the transition is slightly smoothed out by the finite value of $\omega_{0}$ used in the simulations, as discussed above.

\begin{figure}[ht]
\centerline{\includegraphics[width=0.4\textwidth]{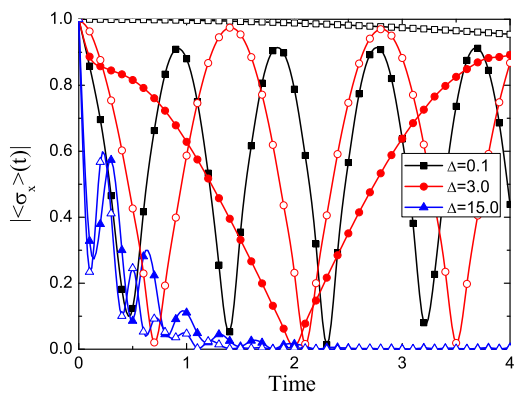}}
\caption{ (Color online) Time evolution of the absolute values of the coherence $\langle\sigma_x(t)\rangle$  for RWA (filled plot markers) and the full Hamiltonian (empty plot markers). For these data $\alpha=1$,$\omega_{0}=100$, $\omega_{b}=5$, and $\omega_{c}=800$. Time is in units of $1/\alpha^2$}
\label{fig4}
\end{figure}

\begin{figure}[ht]
\centerline{\includegraphics[width=0.39\textwidth]{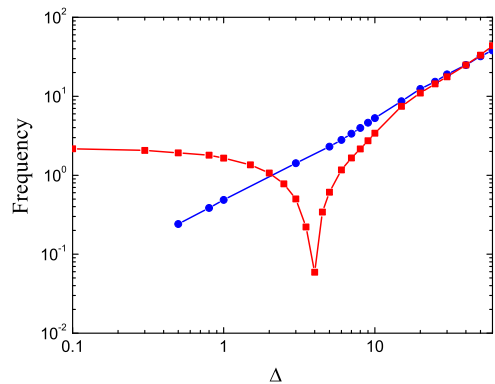}}
\caption{ (Color online) Frequency of coherence oscillations using the RWA (red squares) and the complete Hamiltonian (blue dots) as a function of $\Delta$ for an initially-prepared superposition of ground and excited states. Parameters are $\alpha=1$, $\omega_{0}=100$, $\omega_{b}=5$, and $\omega_{c}=800$. }
\label{fig5}
\end{figure}

\section{Coherence dynamics and the failure of the RWA}\label{coherences}
Let us now consider the evolution of the system coherences $\langle\sigma_x(t)\rangle$, starting from an initial condition where $\langle\sigma_x(0)\rangle=1$. In the absence of the radiation field, we would expect that $\langle\sigma_x(t)\rangle=\cos(\Delta t)$ and in its  presence these coherent oscillations would - in a simple markovian picture - become damped and shifted to lower frequencies \cite{breuerbook, leggett1987}. Figure \ref{fig4} shows the dynamics of $|\langle\sigma_x(t)\rangle|$. The difference in dynamical behaviour is dramatic. As $\Delta$ is increased from well below the gap to well above, the full simulation shows monotonically increasing frequency for the coherence oscillations, with some very weak non-markovian dephasing just below the band edge and much faster, exponential decay for $\Delta=15$. In stark contrast, the RWA oscillations show a \emph{non-monotonic} dependence on $\Delta$, initally decreasing in frequency and then increasing as $\Delta$ increases.

Figure \ref{fig5} shows the frequency of the coherence oscillations for the full Hamiltonian and the RWA for different values of the atomic frequency $\Delta$. The behaviour seen for the RWA curve in Figs. \ref{fig4} and \ref{fig5} can be understood by looking at the pole structure of the exact RWA solution expressed as a Laplace transform. For our initial condition, the Laplace transform of $\langle\sigma_{x}(t)\rangle$ can be expressed as $$\mathcal{L}[\langle \sigma_{x}(t)\rangle]=\frac{1}{2}(s+i\Delta+\mathcal{L}[G(t)])^{-1}+h.c.$$
The poles of $\mathcal{L}[\langle \sigma_{x}(t)\rangle]$ determine the dynamics of the coherence, with their imaginary parts giving the oscillation frequency and their real parts quantifying the dephasing rate. Although the Laplace transform $\mathcal{L}[G(t)]$ can be computed exactly, we can qualitatively understand the key features of Fig. \ref{fig5} by considering just an expansion to lowest order in the small quantities $\Delta/\omega_{0},\omega_{b}/\omega_{0}$. This gives $$\mathcal{L}[G(t)])=-i\alpha\sqrt{\frac{\omega_{0}}{\pi}}+\alpha\sqrt{is-\omega_{b}},$$ with $\alpha$ given in Eq. (\ref{spectralgap}). When $\Delta\rightarrow 0$, the poles of $\mathcal{L}[\langle \sigma_{x}(t)\rangle]$, $s_{\pm}$, are approximately $$s_{\pm}\approx\pm i\alpha(\sqrt{\frac{\omega_{0}}{\pi}}-\alpha\sqrt{\omega_{b}}).$$ This shows that the oscillation frequency converges to a finite value as $\Delta\rightarrow 0$, as is observed in Fig. \ref{fig5}. The poles are completely imaginary, and these oscillations are undamped, although there is a slight loss of amplitude due to dynamical transients. For small but finite $\Delta$, the poles now appear at approximately $$s_{\pm}\approx \pm i(\Delta-\alpha\sqrt{\frac{\omega_{0}}{\pi}}+\alpha\sqrt{\Delta-\omega_{b}}),$$ leading to a reduction of the effective oscillation frequency as the $\Delta$ term compensates the frequency shift coming from the environment. The large dip in oscillation frequency seen in Fig. \ref{fig5} corresponds to a significant cancellation between these terms.  For large values of $\Delta$ ($\Delta\gg \omega_b$), the poles are dominated by the contribution from $\Delta$ and are approximately $$s_{\pm}\approx\pm i\Delta+\alpha\sqrt{\Delta}.$$ The oscillation frequency is now proportional to $\Delta$ and has acquired a real part which damps the oscillations with a rate given by the Fermi Golden Rule as $\Gamma=J(\Delta)$. This is the result one would obtain by a standard weak coupling and Markovian approximation to the open-system dynamics with the spectral density of Eq. (\ref{spectralgap}).

The behaviour of the oscillation frequency for small $\Delta$ is clearly a pathology of the RWA approximation and arises because of a poor treatment of the low frequency modes in the problem. The breakdown of the RWA manifests itself in the renormalisation of the atomic energy $\Delta$ which leads to a shift in $\Delta$ that is larger than $\Delta$ itself, an effect which produces the minimum in Fig. \ref{fig5} and the convergence to a finite oscillation frequency as $\Delta\rightarrow 0$. The correct behaviour in this limit can be calculated exactly for the full Hamiltonian (\ref{genH2}), as for $\Delta=0$ the model corresponds to the exactly solvable independent boson model \cite{mahan1981}. This shows, as one would expect, that $\langle \sigma_{x}(t)\rangle$ does not oscillate and remains constant at $\Delta=0$. This limiting behaviour is correctly described by the TEDOPA simulation of the full Hamiltonian. For small, but finite $\Delta$, the frequency simply increases monotonically with $\Delta$ and the TEDOPA results agree very well with the prediction $$\tilde{\Delta}\approx\Delta\exp(-\alpha/\sqrt{\omega_{b}})+O(\frac{\omega_{b}}{\omega_{0}})$$ given by adiabatic renormalisation theory (a simplified version of Silbey-Harris theory which is valid in the small $\Delta$ limit) \cite{leggett1987}.

We finally remark that the failure of the RWA coherences also manifests itself in the frequencies of the populations we presented in Fig. \ref{fig2}. Neglecting dissipative effects, the oscillation frequency in the excited state population below the band gap would be expected to vary as $\Delta_{L}= (\tilde{\Delta}-\omega_{b})$, and would thus vanish as the \emph{renormalised} atomic energy moves into the bandwidth of the light field. By comparison with Fig. \ref{fig5}, it can be seen that for the values of $\Delta$ considered in Fig. \ref{fig2}, the over-suppression of $\tilde{\Delta}$ for the RWA leads to both a suppression of population decay (atomic transition is further away from the band) and, consequently, spuriously fast population oscillations. This is best seen in Fig. \ref{fig2} at $\Delta=\omega_{b}=5$, where the oscillatory part of the RWA population dynamics is almost four times faster than in the full Hamiltonian simulation and retains $50\%$ more of the excited state population. This failure of the RWA renormalisation for low values of $\Delta$ also leads to artificially large residual population in the excited (relative to the full simulation) seen in Fig. \ref{fig3} for $\Delta=<11$.

\section{Application to photonic crystals and strongly-interacting light matter systems}
Considering a photonic crystal with a gap in the optical region, couplings of the order of $\alpha^2=10^5-10^6 \textit{Hz}$ may be achieved, which in our units \cite{units} would be $\alpha^2=10^{-9}-10^{-10}$. In order to enlarge the coupling, and thus reach the regime discussed in this paper, there are different possibilities. One may consider for instance photonic crystals with narrow band-widths $\Delta\omega$, what would give rise to smaller cut-off frequencies $\omega_0=\omega_{k_0}=\omega_b+l \Delta\omega  k_0^2/3$. Here we have considered the fact that the $\hbar/2m=l^2\Delta\omega/3$ \cite{devega2005}.

Strong system-environment coupling can also be achieved when strong collective effects produce an enhanced effective system-environment coupling \cite{dicke1954,cooperative}. The simplest situation in which this enhancement can be achieved is by considering a collection of $N$ atoms uniformly interacting with the radiation field or environment, or similarly, when describing the collective coupling of a spin system to superconducting resonators \cite{kubo2010}.
In this case, for a single excitation in the system (i.e. only one atom in the ensemble is excited), the situation is completely analogous to the one described here, except for the fact that the coupling is enhanced by a factor $N^2$ in the photonic crystal case \cite{john1995}.  Thus, a value $\alpha^2_{\eeff}=N^2\alpha^2=1$ would be achieved if we have a high enough number of impurity atoms. Of particular interest is also the multi-excitation regime for this configuration.  A spontaneous polarization phenomenon, in which the atomic coherences grows from zero to a finite value, is predicted to occur for atoms in photonic crystal-like environments, when the atomic frequencies are resonant with the band edge $\Delta=\omega_b$ and when considering the semi-classical and RW approximations \cite{john1995,devega2008,navarrete2010}.  The results in our work show strong differences in the behaviour of single atom coherences with and without considering the RWA, particularly for small $\Delta_L$, and suggest that applying the novel technique \cite{prior2010,chin2010} for treating the system-environment interaction in the analysis of multi-atomic systems will shed new light into this and other intriguing collective phenomena.
Other possible enhancement factors for the coupling coefficient would involve placing the atom within a photonic crystal cavity \cite{pcbook}. In a more general scenario, one may consider a situation where the effective system-environment interaction, characterised by $\sqrt{\eta_{0}}$ in our chain representation of the system, is larger than the typical frequencies of the light field, $\omega_{b}$. As we have shown, in this regime the ground and excited atomic states are entangled with the light field through polaronic correlations, and these are very similar to the shifted vacuum states which appear in the recent theory of \emph{ultrastrong coupling} in quantum wells and circuit QED where this regime can be realised \cite{ciuti2010,ciuti2011}.
In conclusion, this work has demonstrated that our TEDOPA method for the treatment of system environment interaction has allowed us to reveal strong deviations between the exact and approximate solutions in physical models of practical importance,  and hence suggests its use in a wide variety of settings where non-perturbative system environment interactions are expected to play a role.

We thank A. Imamoglu for his comments at the start of this project, the EU-STREP PICC and the EU Integrated project QESSENCE, as well as the Alexander von Humboldt foundation for support. J.P. was supported by the Fundaci{\'o}n S\'eneca Project No. 11920/PI/09-j and the Ministerio de Econom{\'i}a y Competitividad Project No. FIS2012-30625 and HOPE-CSD2007-00007 (Consolider Ingenio), I.d.V was partially supported by the Ministerio de Ciencia e Innovaci{\'o}n Project No. FIS2010-19998. A.W.C. acknowledges support from the Winton Programme for the Physics of Sustainability.


\begin{thebibliography}{10}
\bibitem{shift}
In these inequalities, the shift $\omega_s$ has been neglected.  .
\end{thebibliography}
\end{document}